\documentclass{article}
\usepackage{spconf,amsmath,graphicx,amsfonts,epsfig}
\usepackage{times}
\usepackage{amsbsy}
\usepackage{latexsym}
\usepackage{amssymb}

\title{Adaptive Link Selection Strategies for Distributed Estimation in Wireless Sensor Networks}

\name{}
\address{}
\name{Songcen Xu, Rodrigo C. de Lamare and H. Vincent Poor}
\address{Communications Research Group, Department of Electronics\\
University of York, U.K.\\
Princeton University \\
Emails: sx520@ohm.york.ac.uk, rcdl500@ohm.york.ac.uk,
poor@princeton.edu}

\begin{document}
\ninept

\linespread{0.975}
\maketitle
\begin{abstract}

In this work, we propose adaptive link selection strategies for
distributed estimation in diffusion-type wireless networks. We
develop an exhaustive search-based link selection algorithm and a
sparsity-inspired link selection algorithm that can exploit the
topology of networks with poor-quality links. In the exhaustive
search-based algorithm, we choose the set of neighbors that results
in the smallest excess mean square error (EMSE) for a specific node.
In the sparsity-inspired link selection algorithm, a convex
regularization is introduced to devise a sparsity-inspired link
selection algorithm. The proposed algorithms have the ability to
equip diffusion-type wireless networks and to significantly improve
their performance. Simulation results illustrate that the proposed
algorithms have lower EMSE values, a better convergence rate and
significantly improve the network performance when compared with
existing methods.
\end{abstract}
\begin{keywords}
Adaptive link selection, diffusion networks, wireless sensor
networks, distributed processing.
\end{keywords}
\section{Introduction}
\label{sec:intro}

Distributed strategies have become very popular for parameter
estimation in wireless networks and applications such as sensor
networks and smart grids \cite{Lopes1,Lopes2,Chen,Xie}. In this
context, for each specific node a set of neighbor nodes collect
their local information and transmit the estimates back to a
specific node. Many works in the literature have proposed strategies
for distributed processing: incremental \cite{Lopes1}, diffusion
\cite{Lopes2}, sparsity-aware \cite{Chen} and consensus-based
strategies \cite{Xie}. With diffusion strategies \cite{Lopes2} and a
least-mean square (LMS) algorithm, the neighbors for each node are
fixed and the combined coefficients are pre-calculated after the
network is generated. This approach may not provide an optimized
estimation performance for each specified node because there are
links that are more severely affected by noise or fading. Moreover,
when the number of neighbor nodes is large, each node requires a
large network bandwidth and transmit power. In order to reduce this
bandwidth requirement, a single-link diffusion strategy has been
reported recently in \cite{Zhao}, where each node selects its best
neighbor node. A key problem with the strategies reported so far in
the literature is that they do not exploit the topology of wireless
networks and the knowledge about the poor links to improve the
performance of estimation algorithms.


In order to optimize the performance of distributed estimation
techniques in wireless networks and minimize the excess mean-square
error (EMSE) associated with the estimates, we propose two adaptive
link selection algorithms. The proposed techniques exploit the
knowledge about the poor links and the topology of the network in
order to select a subset of links that result in an improved
estimation performance. The first approach consists of an exhaustive
search–based link selection (ESLS) algorithm, whereas the second
technique is based on a sparsity-inspired link selection (SILS)
algorithm. For the ESLS algorithm, we consider all possible
combinations for each node with its neighbors and choose the
combination associated with the smallest EMSE value. For the SILS
algorithm, we introduce the RZA strategy into the adaptive link
selection algorithm. The RZA strategy is usually employed in
applications to deal with sparse systems in such a way that it
shrinks the small values in the weight vector to zero, which results
in a better convergence rate and steady-state performance. Unlike
prior work with sparsity-aware algorithms
\cite{Chen,Rcdl1,Rcdl2,Meng,Guo,zhaocheng12}, the proposed SILS
algorithm exploits the possible sparsity of the EMSE associated with
each of the links in an opposite way. We introduce a convex penalty,
i.e., an $\ell_1$--norm term to adjust the combined coefficients for
each node with its neighbors in order to select the neighbor nodes
that yield the smallest EMSE values. For a specified node, we
calculate the EMSE all its neighbor nodes including the specified
node itself through the previous estimation. For the node with the
maximum EMSE, we impose a penalty and give the same amount of award
to the node with the minimum EMSE. The proposed SILS algorithm
performs this process automatically. By using the SILS algorithm
some nodes with bad performance will be shrunk and some poor nodes
will taken into account when their performance improves, which means
the network topology will change automatically as well. Simulation
results for an application to distributed system identification
illustrate that, the proposed ESLS and SILS algorithms have a better
convergence rate and lower EMSE value when compared with the
existing diffusion LMS strategies \cite{Lopes2}.

This paper is organized as follows. Section 2 describes the
distributed processing in the networks and problem statement. In
section 3, the proposed link selection algorithms are introduced.
The numerical simulation results are provide in section 4. Finally,
we conclude the paper in section 5.

Notation: We use boldface upper case letters to denote matrices and
boldface lower case letters to denote vectors. We use $(\cdot)^T$
and $(\cdot)^{-1}$ denote the transpose and inverse operators
respectively, and $(\cdot)^*$ for conjugate transposition.

\section{Distributed Processing in Wireless Networks and Problem Statement}
\label{sec:prob}

\begin{figure}[!htb]
\begin{center}
\def\epsfsize#1#2{1.0\columnwidth}
\epsfbox{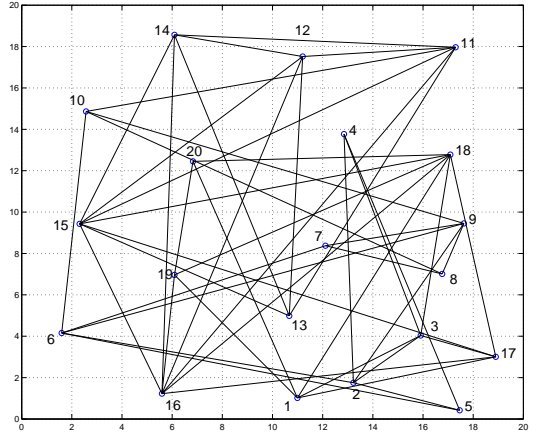} \vspace{-1.85em} \caption{\footnotesize
Network topology with 20 nodes}\vspace{-0.5em}
\label{fig1}
\end{center}
\end{figure}

We consider a diffusion--type wireless network with N nodes which
have limited processing capabilities. At every time instant $i$,
each node $k$ takes a scalar measurement $d_k^{(i)}$ according to:
\begin{equation}
{d_k^{(i)}} = {\boldsymbol {\omega}}_0^H{\boldsymbol x_k^{(i)}} +{n_k^{(i)}},~~~
i=1,2, \ldots, N ,
\end{equation}
where ${\boldsymbol x_k^{(i)}}$ is the $M \times 1$ input signal
vector, ${ n_k^{(i)}}$ is the noise sample at each node with zero
mean and variance $\sigma_{v,k}^2$. Through (1), we can see that the
measurements for all nodes are related to an unknown vector
${\boldsymbol {\omega}}_0$. Fig.\ref{fig1} shows an example for a
diffusion -- type wireless network with 20 nodes. The aim for the
diffusion -- type network is to compute an estimate of ${\boldsymbol
{\omega}}_0$ in a distributed fashion, which can minimize the cost
function:
\begin{equation}
{J_{\omega}({\boldsymbol \omega})} = {\mathbb{E} |{ d_k^{(i)}}-
{\boldsymbol \omega}^H{\boldsymbol x_k^{(i)}}|^2} ,
\end{equation}
where $\mathbb{E}$ denotes the expectation operator. To solve this
problem, one possible basic diffusion strategy is the
adapt--then--combine (ATC) diffusion strategy \cite{Lopes2}:
\begin{equation}
\left\{\begin{array}{ll}
{\boldsymbol {\psi}}_k^{(i)}= {\boldsymbol {\omega}}_k^{(i-1)}+{\mu}_k {\boldsymbol x_k^{(i)}}[{ d_k^{(i)}}-
{\boldsymbol \omega}_k^{(i-1)*}{\boldsymbol x_k^{(i)}}]^*,\\
{\boldsymbol {\omega}}_k^{(i)}= \sum\limits_{l\in \mathcal{N}_k} c_{kl} \boldsymbol\psi_l^{(i)},
\end{array}
\right.
\end{equation}
where $c_{kl}$ is the combine coefficient, which is calculated under the Metropolis rules:
\begin{equation}
\left\{\begin{array}{ll}
c_{kl}= \frac{1}
{max(n_k,n_l)},\ \ \ \ \ \ \ \ \  \ \ \ \ \ \ \ \ \ \ \ \ \ \ \ \ \ \ \     $if\  $k\neq l$\  \ are\  linked$\\
c_{kl}=0,              \ \ \ \ \ \ \ \ \ \  \ \ \ \ \ \ \ \ \ \ \ \ \ \ \ \ \ \ \ \ \ \ \ \ \ \ \ \ \ \ \ \ \ \     $for\  $k$\  and\  $l$\ not\  linked$\\
c_{kk} = 1 - \sum\limits_{l\in \mathcal{N}_k / k} c_{kl}, \ \ \ \ \ \ \ \ \ \ \ \ \ \ \ \ \ \ \ \ \ $for\  $k$\ =\ $l$$
\end{array}
\right.
\end{equation}
and should satisfy:
\begin{equation}
\sum\limits_{l} c_{kl} =1 , l\in \mathcal{N}_k \forall k
\end{equation}
For this kind of strategy, the choice of the neighbors for each node
is fixed, this situation will cause a problem when some of the
neighbor nodes have a poor performance, and there is no chance for
the node to discard the bad performance neighbors instead of
continue to use their information. To optimize the distributed
processing and improve the performance, we need to provide each node
some refinements and the ability to select its links.

\section{Proposed adaptive link selection algorithms}
\label{sec:plsa}

In order to optimize the distributed processing and improve the
performance of the network, we propose the ESLS and the SILS
algorithms. These two algorithmic strategies give the nodes the
ability to choose their neighbors based on their EMSE performance.

\subsection{Exhaustive search-based link selection (ESLS)}
\label{ssec:esls}

The ESLS employs an exhaustive search to select the links that yield
the best performance in terms of EMSE. For our proposed ESLS
algorithm, we employ the adaptation strategy given by
\begin{equation}
{\boldsymbol {\psi}}_k^{(i)}= {\boldsymbol {\omega}}_k^{(i-1)}+{\mu}_k {\boldsymbol x_k^{(i)}}[{ d_k^{(i)}}-
{\boldsymbol \omega}_k^{(i-1)*}{\boldsymbol x_k^{(i)}}]^* \notag
\end{equation}
and redefine the diffusion step.

First, we introduce a tentative set $\Omega_s$ using a combinatorial
approach described by
\begin{equation}
{\Omega_s}\triangleq {C_{nk}^j},~~~j=1,2, \ldots, nk,
\end{equation}
where \{$nk$\} is the total number of nodes linked to node $k$
including node $k$ itself. This combinatorial strategy will cover
all combination choices for each node k with its neighbors.

After the tentative set $\Omega_s$ is defined, we redefine the cost
function as
\begin{equation}
{J_{\psi}({\boldsymbol \psi})} \triangleq {\mathbb{E} |{ d_k^{(i)}}-
{\boldsymbol x_k^{(i)}}^*{\boldsymbol \psi}|^2} ,
\end{equation}
where
\begin{equation}
{\boldsymbol \psi} \triangleq \sum\limits_{l\in \Omega_s} c_{kl} \boldsymbol\psi_l^{(i)},
\end{equation}
and the error pattern is introduced as:
\begin{equation}
{e_{\Omega_s}^{(i)}} \triangleq { d_k^{(i)}}-{\boldsymbol x_k^{(i)}}^*{\sum\limits_{l\in \Omega_s} c_{kl} \boldsymbol\psi_l^{(i)}} .
\end{equation}
For each node $k$, the strategy that finds the best set $\Omega_s$
should solve the following optimization:
\begin{equation}
\widehat{\Omega_s}=\arg\min\limits_{\Omega_s}{J_{\psi}({\boldsymbol \psi})}
\end{equation}
which is equivalent to minimizing the error $e_{\Omega_s}^{(i)}$.
After all steps have been completed, the diffusion step in (3) can
be modified as:
\begin{equation}
{\boldsymbol {\omega}}_k^{(i)}= \sum\limits_{l\in \widehat{\Omega_s}} c_{kl} \boldsymbol\psi_l^{(i)}.
\end{equation}
At this point, the main steps of the ESLS algorithm have been
completed. The proposed ESLS algorithm finds the best set
$\widehat{\Omega_s}$ which minimizes the cost function in (7) and
then uses this set of nodes to obtain ${\boldsymbol
{\omega}}_k^{(i)}$ through (11). The ESLS algorithm is summarized in
Table \ref{table1}. When the ESLS is implemented in networks with
small and low-power sensors, the cost may become expensive, as the
algorithm in (6) requires an exhaustive search and needs more
communication resources to examine all the possible sets $\Omega_s$.

\begin{table}
\centering \caption{The ESLS Algorithm}
\begin{tabular}{l}
\hline
Initialize: ${\boldsymbol {\omega}}_k^{(-1)}$=0 \\
For each time instant $i$=1,2, . . . , n\\
For each node $k$=1,2, \ldots, N\\
\ \ \ \ \ \ \ \ \ \ ${\boldsymbol {\psi}}_k^{(i)}= {\boldsymbol {\omega}}_k^{(i-1)}+{\mu}_k {\boldsymbol x_k^{(i)}}[{ d_k^{(i)}}-
{\boldsymbol \omega}_k^{(i-1)*}{\boldsymbol x_k^{(i)}}]^*$\\
end\\
For each node $k$=1,2, \ldots, N\\
\ \ \ \ \ \ \ \ \ \ find all possible sets of ${\Omega_s}$\\
\ \ \ \ \ \ \ \ \ \ ${e_{\Omega_s}^{(i)}} = { d_k^{(i)}}-{\boldsymbol x_k^{(i)}}^*{\sum\limits_{l\in \Omega_s} c_{kl} \boldsymbol\psi_l^{(i)}} $\\
\ \ \ \ \ \ \ \ \ \ $\widehat{\Omega_s}=\arg\min\limits_{\Omega_s}{e_{\Omega_s}^{(i)}}$\\
\ \ \ \ \ \ \ \ \ \ ${\boldsymbol {\omega}}_k^{(i)}= \sum\limits_{l\in \widehat{\Omega_s}} c_{kl} \boldsymbol\psi_l^{(i)}$\\
end\\
end\\
\hline
\end{tabular}
\label{table1}
\end{table}

\subsection{Sparsity-inspired link selection (SILS)}
\label{ssec:sils}

The ESLS algorithm outlined above mentioned problem needs to examine
all possible sets to find a solution, which might result in an
unacceptable computational complexity for large networks operating
in time-varying scenarios. To solve the combinatorial problem with a
low complexity, we propose the SILS algorithm which is as simple as
a standard diffusion LMS algorithm and is suitable for adaptive
implementations and scenarios where the parameters to be estimated
are slowly time-varying.

\begin{figure}[!htb]
\begin{center}
\def\epsfsize#1#2{1.0\columnwidth}
\epsfbox{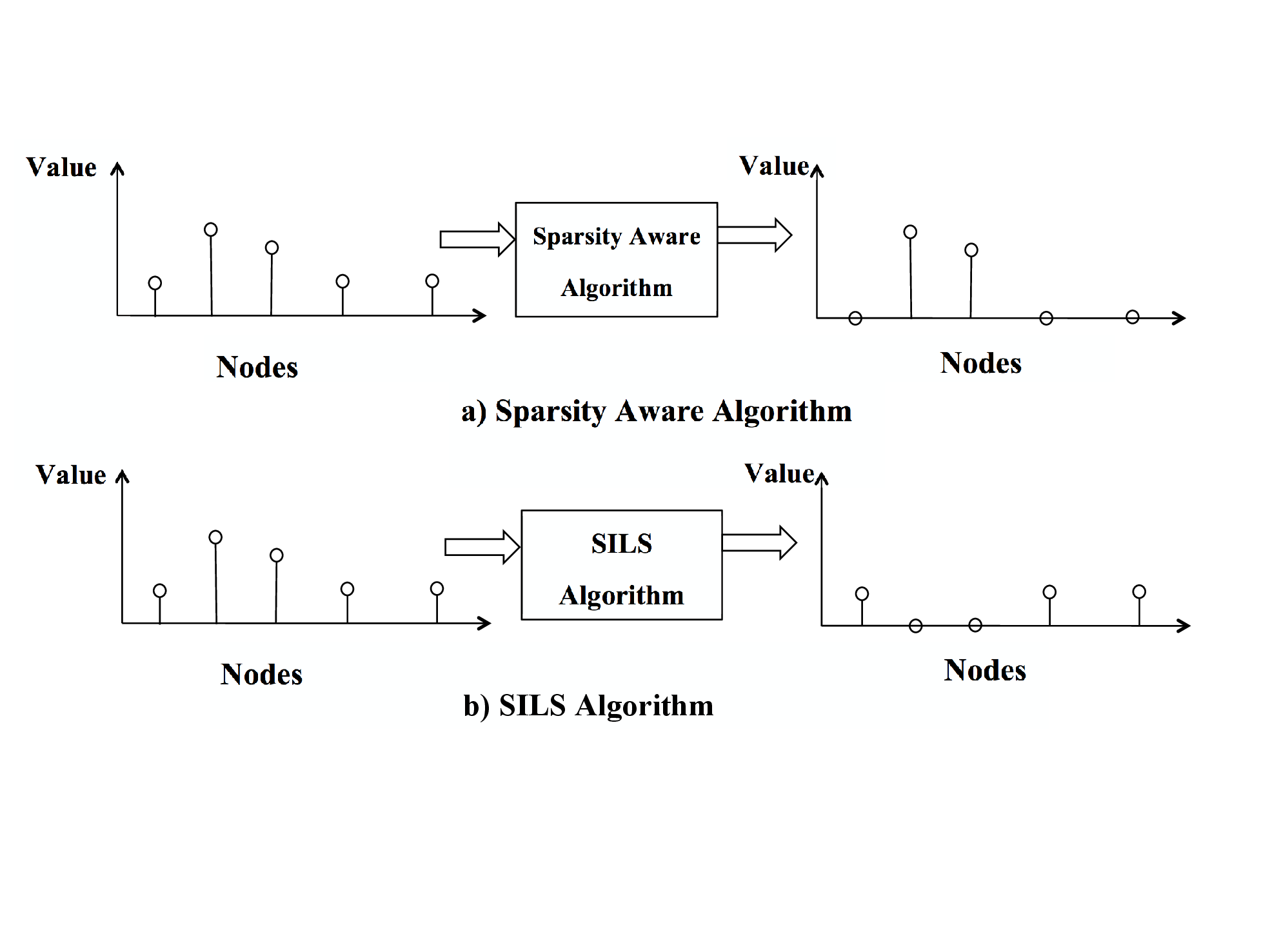} \vspace{-1.85em} \caption{\footnotesize
Sparsity Aware Technology}\vspace{-0.5em}
\label{fig3}
\end{center}
\end{figure}

Fig. 2 shows the difference between the existing sparsity-aware
methods and the proposed sparsity-inspired strategy. In Fig. 2 (a),
we can see that, after being processed by a sparsity-aware
algorithm, the nodes with small error values will be shrunk to zero.
In contrast, the sparsity-inspired algorithm will shrink the nodes
with large error values to zero as illustrated in Fig. 2.

In the proposed SILS algorithm, we introduce the convex penalty term
$\ell_1$--norm into the diffusion step in (3) to perform the link
selection. Different penalty terms have been considered for this
task. We have adopted the reweighted zero--attracting strategy
\cite{Chen} into the diffusion step in (3) because this strategy has
shown an excellent performance and is simple to implement.

First, we consider the following regularization function:
\begin{equation}
{f_1(e_l^{(i)}))}= \sum\limits_{l\in \mathcal{N}_k} \log(1+\varepsilon|e_l^{(i)}|)
\end{equation}
where the error pattern $e_l^{(i)}$ is defined as:
\begin{equation}
{e_l^{(i)}} \triangleq { d_k^{(i)}}-{\boldsymbol x_k^{(i)}}^*{ \boldsymbol\psi_l^{(i)}} \ \ \ \ \ \ \ \ \ (l\in \mathcal{N}_k)
\end{equation}
and $\varepsilon$ is the shrinkage magnitude. Then ,the diffusion
step in (3) can be transformed into the link selection method as:
\begin{equation}
{\boldsymbol {\omega}}_k^{(i)}= \sum\limits_{l\in \mathcal{N}_k} [c_{kl}-\rho \frac{\partial f_1}{\partial e_l}(e_l^{(i)})] \boldsymbol\psi_l^{(i)}
\end{equation}
where $\rho$ is used to control the strength of the algorithm and it
is updated by using
\begin{equation}
\frac{\partial{f_1(e_l^{(i)})}}{\partial e_l}=
\varepsilon\frac{{\rm{sign}}(e_l^{(i)}) }{1+\varepsilon|\xi_{\rm
min}|}
\end{equation}
In (15), the parameter $\xi_{\rm min}$ stands for the minimum value
of $e_l^{(i)}$ in each group of nodes including each node $k$ and
its neighbors. The function ${\rm{sign}}(x)$ is defined as
\begin{equation}
{{\rm{sign}}(x)}=
\left\{\begin{array}{ll}
{x/|x|}\ \ \ \ \ {x\neq 0}\\
0\ \ \ \ \ \ \ \ \ \ \ \ {x= 0}
\end{array}
\right.
\end{equation}
To further simplify the expression in (14), we introduce the vector
and matrix quantities required to describe the adaptation process.
We first define a vector $\boldsymbol c$ that contains the
combination coefficients for each group of nodes including node $k$
and its neighbors as described by
\begin{equation} {\boldsymbol
c}\triangleq[c_{(k,l)}]_{l\in \mathcal{N}_k}
\end{equation}
Then, we introduce a matrix $\boldsymbol \Psi$ that includes all the
estimated vectors which are generated after the adaptation step in
(3) for each group as given by
\begin{equation}
{\boldsymbol \Psi}\triangleq[\psi_l^{(i)}]_{l\in \mathcal{N}_k}
\end{equation}
An error vector $\boldsymbol e$ that contains all the error values
calculated through (13) for each group is expressed by
\begin{equation}
{\boldsymbol e}\triangleq[e_l^{(i)}]_{l\in \mathcal{N}_k}
\end{equation}
To employ the sparsity --inspired approach, we have modified the
vector $\boldsymbol e$ in the following way. The maximum value
$e_l^{(i)}$ in $\boldsymbol e$ will be set to $|e_l^{(i)}|$, while
the minimum value $e_l^{(i)}$ will be set to $-|e_l^{(i)}|$. For the
remaining entries, they will be set to zero. Finally, by inserting
(15)-(19) into (14), the diffusion step will be changed to
\begin{equation}
\begin{split}
{\boldsymbol {\omega}}_k^{(i)}&={\sum\limits_{j=1}^{\mathcal{N}_k} [\boldsymbol c_j-\rho \frac {\partial f_1}{\partial \boldsymbol e_j}(\boldsymbol e_j)] \boldsymbol\Psi_j}\\
&={\sum\limits_{j=1}^{\mathcal{N}_k} [\boldsymbol c_j-\rho
\varepsilon\frac{{\rm{sign}}({\boldsymbol e_j})
}{1+\varepsilon|\xi_{\rm min}|}] \boldsymbol\Psi_j}
\end{split}
\end{equation}
The proposed SILS algorithm performs link selection by the
adjustment of the combination coefficients through $\boldsymbol c$
in (20). For the neighbor node with the largest EMSE value, after
our modifications for $\boldsymbol e$, its $e_l^{(i)}$ value in
$\boldsymbol e$ will be a positive number which will lead to the
term $\rho\varepsilon\frac{{\rm{sign}}({\boldsymbol
e_j})}{1+\varepsilon|\xi_{\rm min}|}$ in (20) being positive too.
This means that the combining coefficient for this node will be
reduced and the weight for this node to build the ${\boldsymbol
{\omega}}_k^{(i)}$ is reduced too. In contrast, for the neighbor
node with the minimum EMSE, as its $e_l^{(i)}$ value in $\boldsymbol
e$ will be a negative number, the term
$\rho\varepsilon\frac{{\rm{sign}}({\boldsymbol
e_j})}{1+\varepsilon|\xi_{\rm min}|}$ in (20) will be negative too.
As a result, the weight for this node associated with the minimum
EMSE to build the ${\boldsymbol {\omega}}_k^{(i)}$ is increased. For
the remaining neighbor nodes, the $e_l^{(i)}$ value in $\boldsymbol
e$ is zero, which means the term
$\rho\varepsilon\frac{{\rm{sign}}({\boldsymbol
e_j})}{1+\varepsilon|\xi_{\rm min}|}$ in (20) is zero and there is
no change for their weights to build the ${\boldsymbol
{\omega}}_k^{(i)}$. The process for the combination coefficients is
still satisfied (5). The SILS algorithm is summarized in Table
\ref{table2}.

For the ESLS and SILS algorithms, we redesign the diffusion step and
employ the same adaptation procedure, which means these two
algorithms have the ability to equip any diffusion -- type wireless
networks besides the LMS strategy. This includes the diffusion RLS
strategy \cite{Cattivelli} and the diffusion conjugate gradient
strategy \cite{Xu}.

\begin{table}
\centering \caption{The SILS Algorithm}
\begin{tabular}{l}
\hline
Initialize: ${\boldsymbol {\omega}}_k^{(-1)}$=0 \\
For each time instant $i$=1,2, . . . , n\\
For each node $k$=1,2, \ldots, N\\
\ \ \ \ \ \ \ \ \ \ ${\boldsymbol {\psi}}_k^{(i)}= {\boldsymbol {\omega}}_k^{(i-1)}+{\mu}_k {\boldsymbol x_k^{(i)}}[{ d_k^{(i)}}-
\grave{}{\boldsymbol \omega}_k^{(i-1)*}{\boldsymbol x_k^{(i)}}]^*$\\
end\\
For each node $k$=1,2, \ldots, N\\
\ \ \ \ \ \ \ \ \ \ ${e_l^{(i)}} = { d_k^{(i)}}-{\boldsymbol x_k^{(i)}}^*{ \boldsymbol\psi_l^{(i)}} \ \ \ \ \ \ \ \ \ (l\in \mathcal{N}_k)$\\
\ \ \ \ \ \ \ \ \ \ ${\boldsymbol c}=[c_{(k,l)}]_{l\in \mathcal{N}_k}$\\
\ \ \ \ \ \ \ \ \ \ ${\boldsymbol \Psi}=[\psi_l^{(i)}]_{l\in \mathcal{N}_k}$\\
\ \ \ \ \ \ \ \ \ \ ${\boldsymbol e}=[e_l^{(i)}]_{l\in \mathcal{N}_k}$\\
\ \ \ \ \ \ \ \ \ \ Find the maximum and minimum terms in ${\boldsymbol e}$\\
\ \ \ \ \ \ \ \ \ \ Modified ${\boldsymbol e}$ as ${\boldsymbol e}$=[0$\cdot\cdot\cdot$0,$\underbrace{|e_l^{(i)}|}_{\max}$,0$\cdot\cdot\cdot$0,$\underbrace{-|e_l^{(i)}|}_{\min}$,0$\cdot\cdot\cdot$0]\\
\ \ \ \ \ \ \ \ \ \ $\xi_{\rm max}= \min(e_l^{(i)})$\\
\ \ \ \ \ \ \ \ \ \ ${\boldsymbol {\omega}}_k^{(i)}= {\sum\limits_{j=1}^{\mathcal{N}_k} [\boldsymbol c_j-\rho \varepsilon\frac{{\rm{sign}}({\boldsymbol e_j}) }{1+\varepsilon|\xi_{\rm min}|}] \boldsymbol\Psi_j}$\\
end\\
end\\
\hline
\end{tabular}
\label{table2}
\end{table}

\section{Simulation Results}
\label{sec:sim}

In this section, we compare our proposed diffusion link selection
algorithms ESLS and SILS with the traditional diffusion ATC
algorithm \cite{Lopes2} based on the performance of EMSE. With the
network topology structure in Fig. \ref{fig1}, we introduce N=20
nodes in this system. The length for the unknown parameter
$\omega_0$ is M=10 and it is generated randomly. The input signal is
generated as ${\boldsymbol u_k^{(i)}}=[u_k^{(i)}\ \ \ u_k^{(i-1)}\ \
\ ...\ \ \ u_k^{(i-M+1)}]$  and
$u_k^{(i)}=x_k^{(i)}+\rho_ku_k^{(i-1)}$, where $x_k^{(i)}$ is a
white noise process to ensure the variance of ${\boldsymbol
u_k^{(i)}}$ $\sigma^2_{u,k}= 1$. The noise samples are modeled as
complex Gaussian noise with variance of $\sigma^2_k= 0.001$. The
step size for all these three algorithms is $\mu=0.045$. For the
static scenario, the sparsity parameters of SILS algorithm are set
to $\rho=4*10^{-3}$ and $\varepsilon=10$. The results are averaged
over 100 independent runs. From Fig. \ref{fig3}, we can see that,
the ESLS has the best performance on both the EMSE and convergence
rate, it has around a 5 dB gain over the traditional diffusion ATC
algorithm. SILS is a bit worse than the ESLS, but still
significantly better than the standard diffusion ATC algorithm by
about 4 dB. For the complexity and processing time, SILS is as
simple as the standard diffusion ATC algorithm, while ESLS is more
complex. For the time -- varying scenario, the sparsity parameters
of SILS algorithm are set to $\rho=6*10^{-3}$ and $\varepsilon=10$.
The unknown vector ${\boldsymbol {\omega}}_0$ is is defined by the
first -- order Markov vector process:
\begin{equation}
{\boldsymbol \omega_0^{(i+1)}}={\boldsymbol \omega_0^{(i)}}+{\boldsymbol z^{(i)}}
\end{equation}
where $\boldsymbol z^{(i)}$ is an independent zero -- mean gaussian vector process.
Fig. \ref{fig4} shows that, for the time -- varying scenario,
the ESLS still performs best, while the SILS has the second best performance.

\begin{figure}[!htb]
\begin{center}
\def\epsfsize#1#2{1.0\columnwidth}
\epsfbox{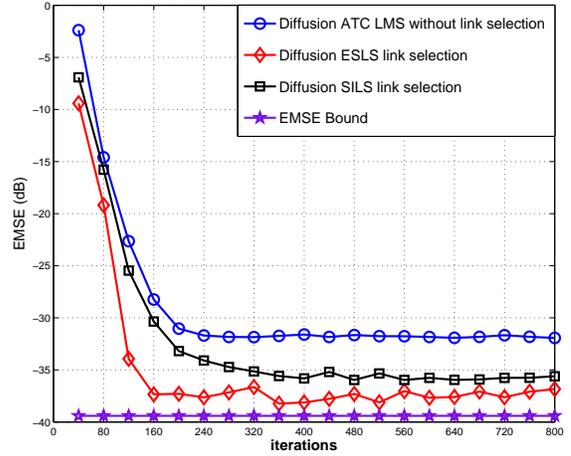} \vspace{-1.85em} \caption{\footnotesize
Network EMSE curves in a static scenario}\vspace{-0.5em}
\label{fig3}
\end{center}
\end{figure}

\begin{figure}[!htb]
\begin{center}
\def\epsfsize#1#2{1.0\columnwidth}
\epsfbox{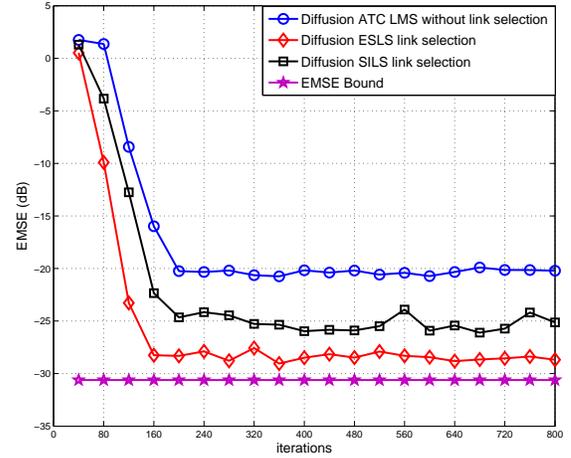} \vspace{-1.85em} \caption{\footnotesize
Network EMSE curves in a time -- varying scenario}\vspace{-0.5em}
\label{fig4}
\end{center}
\end{figure}
\section{Conclusion}
\label{sec:con}

In this paper, two adaptive link selection strategies have been
proposed for distributed estimation in diffusion--type wireless
networks. The ESLS algorithm uses an exhaustive search to perform
the link selection, and the SILS employs a sparsity-inspired
approach with the $\ell_1$ -- norm penalization. The ESLS algorithm
chooses the best set of nodes, while the SILS algorithm shrinks the
node with the highest error values and awards the node with the
smallest errors in each group. Numerical results have shown that the
two proposed algorithms achieve a better convergence rate and lower
EMSE values than the algorithms in \cite{Lopes2}. These results hold
for other algorithms including RLS and CG techniques. 
The ESLS and SILS algorithms can be used in any kind of diffusion --
type wireless networks and applied to problems of statistical
inference in smart grids.

\bibliographystyle{IEEEbib}
\bibliography{reference}

\end{document}